\documentclass{aastex}
\usepackage{natbib}

\shorttitle{Polarimetric noise and sunspot twist}
\shortauthors{Gosain et al.}

\begin{document}

\title{On the estimate of magnetic non-potentiality of sunspots derived using {\it Hinode} SOT/SP observations: Effect of polarimetric noise}
\author{Sanjay Gosain, Sanjiv Kumar Tiwari and P. Venkatakrishnan}
\affil{Udaipur Solar Observatory, Physical Research Laboratory,
Udaipur, Rajasthan- 313 001, INDIA}

\begin{abstract}
The accuracy of Milne-Eddington (ME) inversions, used to
retrieve the magnetic field vector, depends upon the
signal-to-noise ratio (SNR) of the spectro-polarimetric
observations. The SNR in real observations varies from pixel to
pixel, therefore the accuracy of the field vector also varies
over the map. The aim of this work is to study the effect of
polarimetric noise on the inference of magnetic field vector
and the magnetic non-potentiality of a real sunspot. To this
end, we use {\it Hinode} SOT/SP vector magnetogram of a real
sunspot NOAA 10933 as an input to generate synthetic Stokes
profiles under ME model assumptions. We then add
normally-distributed polarimetric noise of the level 0.5\% of
continuum intensity to these synthetic profiles and invert them
again using ME code. This process is repeated 100 times with
different realizations of noise. It is found that within most
of the sunspot area ($>$ 90\% area) the spread in the (i) field
strength is less than 8 Gauss, (ii) field inclination is less
than 1 degree, and (iii) field azimuth is less than 5 degrees.
Further, we determine the uncertainty in the magnetic
non-potentiality  of a sunspot as determined by the force-free
parameter $\alpha_g$ and Spatially Averaged Signed Shear Angle
(SASSA). It is found that for the sunspot studied here these
parameters are $\alpha_g = -3.5 \pm 0.37~ (\times 10^{-9}
m^{-1})$  and SASSA =$ -1.68 \pm 0.014~ ^\circ$. This suggests
that the SASSA is a less dispersion non-potentiality parameter
as compared to $\alpha_g$. Further, we examine the  effect of
increasing noise levels viz. 0.01, 0.1, 0.5 and 1\% of
continuum intensity and find that SASSA is less vulnerable to
noise as compared to $\alpha_g$ parameter.
\end{abstract}

\section{Introduction}
Accurate determination of the vector magnetic field is very
important for the studies of magnetic non-potentiality in
active regions. The evolution of the active region magnetic
field towards an increasingly non-potential state leads to
buildup of magnetic free energy, i.e., energy above magnetic
potential energy. This free magnetic energy is believed to
drive the eruptive phenomena like flares and Coronal Mass
Ejections (CMEs).  Prediction of such phenomena on the Sun is
very important for space weather forecasting, and requires good
knowledge of the non-potentiality  of the magnetic field in
solar active regions.

Force-free parameter $\alpha_g$ has been studied to infer the
 non-potentiality of sunspots for a long time
\citep{Pevtsov94,Pevtsov95,Abramenko96,Bao98,Hagino04,Hagino05,Nandy06}.
 Using the second moment of
minimization \citep{Hagino04,Tiwari09a} the relation between
$\alpha_g$ and the vector field components is given by the
following relation :
\begin{equation}
\alpha_g~=~\frac{\sum(\frac{\partial B_y}{\partial x} - \frac{\partial B_x}{\partial y})B_z}{\sum B_z^2}
\end{equation}
The global alpha ($\alpha_g$) actually gives twice the degree
of twist per unit axial length and not the twist (see Appendix
A of \cite{Tiwari09a}). Thus, if the length of the magnetic
field structure in a volume is given, global twist can be
obtained from alpha for force-free fields.

Another parameter, spatially averaged signed shear angle
(SASSA), henceforth denoted as $\widehat{\Psi}$, was recently
proposed by \cite{Tiwari09b} as a measure of magnetic
non-potentiality in sunspots.  This parameter is the spatial
average of the angle between observed magnetic field and
potential field azimuth. It is derived from the following
relation.
\begin{equation}
SASSA~~~~~or~~~~~ \widehat{\Psi}~=~\langle~{\rm atan}~(\frac{B_{yo}~B_{xp} - B_{yp}~ B_{xo}}{B_{xo} ~ B_{xp} + B_{yo}~B_{yp}})~\rangle
\end{equation}
 where $B_{xo}, B_{yo}$ and $B_{xp}, B_{yp}$ are observed and
potential transverse components of the magnetic field,
respectively. The angled-braces represent the spatial average
taken over all pixels except those below noise (see section
2.3). This parameter thus gives average shear angle on the
photospheric boundary and is independent of the force-free
nature as well as shape of the sunspots \citep{pvk09}. The
high-quality {\it Hinode} data has also allowed these authors
to study the contribution of local alpha values of umbral and
penumbral structures to global alpha value of the sunspots.
Further, \cite{Gosain09} evaluated $\widehat{\Psi}$ of active
region NOAA 10930 during X-class flare of 13 December 2006
using vector magnetograms obtained few hours before and after
the flare  by Spectro-Polarimeter (SP) instrument
\citep{Lites07,Ichimoto08} with Solar Optical Telescope (SOT)
onboard  {\it Hinode} satellite \citep{Kosugi07,Tsuneta08}, and
found that the $\widehat{\Psi}$ decreased after the flare.
However, high-cadence vector magnetograms are needed to follow
the evolution of non-potentiality, characterized by
$\widehat{\Psi}$, during the flare interval. Magnetographs
based on tunable filters like Imaging Vector Magnetograph (IVM)
\citep{Mickey96} and Solar Vector Magnetograph (SVM)
\citep{Gosain04,Gosain06} from ground and recently flown
Helioseismic and Magnetic Imager (HMI) onboard Solar Dynamics
Observatory (SDO) from space \citep{Scherrer02} can provide
such high-cadence magnetograms of active regions.

\cite{Tiwari09a} had earlier evaluated the effect of
polarimetric noise on the magnetic twist parameter $\alpha_g$
and magnetic energy using a synthetic bipole \citep{Low82}.
However, for real sunspots the distribution of magnetic field
vector is not the same as in the case of synthetic bipole. Real
sunspots have umbra and penumbra besides fine structure, as
seen by high resolution ground based and space based {\it
Hinode} observations.  In a recent study by \cite{Su09} using
{\it Hinode} observations it was found that the current and
$\alpha$ distribution in a sunspot is not smooth  but has a
fine mesh-like structure with mixed polarity patches of
$\alpha$ in the umbra and radial spine like pattern with
alternating polarity in penumbra of a sunspot. At the umbra
penumbra boundary they found an incomplete  annular ring with
current and $\alpha$ values of sign opposite to that of global
$\alpha$ i.e., $\alpha_g$. Further, \cite{Tiwari09b} have shown
that the distribution of $\alpha$ in penumbra almost cancels to
zero while in umbra there is a net value which bears same sign
as $\alpha_g$ and has same magnitude as $\alpha_g$.  Further,
the new parameter $\widehat{\Psi}$ that has been introduced
recently by \cite{Tiwari09b} needs to be assessed for
robustness in the presence of uncertainties in the magnetic
field parameters of a real sunspot. This work therefore extends
the results of \cite{Tiwari09a} to the case of a real sunspot
observed by {\it Hinode} and to the new parameter
$\widehat{\Psi}$.

The uncertainties in vector magnetic field obtained by fitting
the observed Stokes profiles with model profiles can arise
mainly due to the following reasons : (i) invalidity of the
model atmosphere used, e.g., ME model cannot fit asymmetric
Stokes profiles which arise due to variation  of physical
parameters within the line forming region,  and (ii)
polarimetric noise, e.g., sensitivity in determining the field
parameters depends on the Signal to Noise Ratio (SNR) of the
Stokes profiles. The former reason has been evaluated by
\cite{Plaza2001} by comparing ME and SIR (Stokes Inversion
based on Response-Functions; \cite{Cobo92}) inversions. SIR
inversions yield variation of physical parameters within the
line forming region, by exploiting the asymmetric nature of
Stokes profiles. They have shown that using ME inverted field
parameters is equivalent to using field parameters averaged
over the line forming region obtained by SIR inversions. So,
neglecting effect of Stokes asymmetry we focused on the errors
in the field measurements due to polarimetric noise.

 The paper
is organized as follows. In section 2 we describe the {\it
Hinode} SOT/SP observations and the method of simulating noisy
Stokes profiles and their inversion. The results are presented
in section 3.  The importance  of the non-potentiality
parameters and the inaccuracy in their determination as a
result of polarimetric noise is discussed in section 4. Section
5 gives the conclusions based on the present study and the
future studies required.

\section{Data and Method}
\subsection{Observational Data}
The high resolution spectro-polarimetric observations were made
by the Spectro-Polarimeter (SP) instrument working with Solar
Optical Telescope (SOT) onboard {\it Hinode} space mission. We
choose a regular isolated sunspot in NOAA 10933 during 05
January 2007, when it was located close to the disk center
($\mu=0.99$)  during 07:20 UT. The sunspot was scanned with
``fast-map" observing mode by the SP instrument. In this mode
the spatial sampling along the slit is 0.32\H{} per pixel and
across the slit is 0.29\H{} per pixel with integration time of
1.6 seconds. The spectral lines used for polarimetric
measurement are Fe I 6301.5 and 6302.5 \AA~ line pair. The
vector magnetic field for this sunspot was derived using
Milne-Eddington code named MERLIN, provided under Community
Spectro-polarimetry Analysis Center (CSAC) \citep{Lites07}. It
is an inversion code based on the least-squares fitting of the
observed Stokes profiles using the Levenberg-Marquardt
algorithm and is quick due to parallel computing. It assumes a
standard Milne-Eddington atmosphere to retrieve the magnetic
field vector, line-of-sight velocity, source function, Doppler
broadening as well as the macro-turbulence and the stray light
filling factor.  For Monte-Carlo estimation of error in
magnetic parameters we use another Milne-Eddington inversion
code named {\it Helix} \citep{Lagg04}.  We compared the
magnetic field parameters retrieved by the two codes after
inverting the SOT/SP dataset used in the present study. We made
difference maps of the retrieved magnetic field parameters
i.e., $\delta M=M_{MERLIN}-M_{HELIX}$, where $M =$  $B$ (field
strength), $\gamma$ (inclination) or  $\psi$ (azimuth). Within
the sunspot, the two codes retrieve identical magnetic field
parameters giving zero mean of $\delta M$ and (1-$\sigma$)
standard-deviation of 50 Gauss, 0.75 degrees and 2.5 degrees in
field-strength, inclination and azimuth angle, respectively.
These values are comparable to the standard errors of inversion
by MERLIN, reported in figure 3.  Standard errors mean that if
same inversion code is repeatedly applied to a given observed
(noisy) profile, it will lead to same model atmosphere within
the error bars given by the standard error estimate, provided
the minimum of the merit function is reached
\citep{Bellot2000}. So, the comparison suggests that the two
codes are identical such that the differences in the output are
of same magnitude as the difference of a particular code with
itself.
The reason for choosing {\it Helix} over MERLIN in our
Monte-Carlo error estimation is the in-built routines for
adding noise to the synthetic Stokes profiles in {\it Helix}
code which helps us save time as we avoid adding noise offline.
The azimuthal ambiguity of the transverse component of the
magnetic field vector was resolved using acute angle method
\citep{Harvey69}.

\subsection{The Monte-Carlo Approach for Determination of Uncertainties}
The uncertainties in the field parameters derived using ME
inversion of observed Stokes profiles, purely as a result of
polarimetric noise present in the observed data, are considered
here. There are two ways of determining the uncertainties:\\
(i)  {\it Standard Errors} : Basically, in the Stokes profile
inversions, a merit function is defined as
$$\chi^2 = \frac{1}{\nu} \sum_{k=1}^4\sum_{i=1}^{M} [I_k^{obs}(\lambda_i)-I_k^{syn}(a,\lambda_i)]^2\frac{w_{ki}^2}{\sigma_{ki}^2}$$
where $k$=1 to 4 represent the four Stokes parameters, $i$= 1
to $M$ represent the number of wavelength samples for observed
and synthetic Stokes profiles, $\nu$ represents the number of
degrees of freedom (i.e., the number of observables minus the
number of free parameters), and the vector $a$ consists  the
parameters of the model, $\sigma_{ki}$ represent the noise in
the observations and $w_{ki}$ represents the weight given to
the data points. This merit function, $\chi^2$, is then
minimized by using non-linear least squares method to obtain
$\chi_{min}$. The standard errors in the model parameters
basically depend upon the curvature of the $\chi^2$ function
near the region of minimum in the parameter space
\citep{Press86}. More details about the estimation of the
standard errors in the model parameters derived by Stokes
inversion is given in \cite{Bellot2000}.

(ii) {\it Monte-Carlo Errors} : In Monte-Carlo method we create
several artificial realizations of the spectro-polarimetric
data. This is done by generating the synthetic Stokes profiles
and adding different realizations of noise to them. These
artificial datasets are then inverted to obtain the magnetic
parameters, leading to a collection of results. The errors are
then determined by finding the spread or standard deviation in
the derived parameters.  It may be noted that the Monte-Carlo
errors have no obvious relationship with the standard errors.

\cite{Plaza2001} has shown that in case of SIR inversion of a
single profile, the standard errors are slightly larger than
the estimates from Monte-Carlo method. However, this depends on
the distribution of field vector, and so the comparison of
errors by the two methods may vary from pixel to pixel.
Further, the Monte-Carlo error estimates depend upon the noise
in the data as well as the noise in the process of inversion.
Thus these are useful to establish the statistical significance
of the results.

We use Monte-Carlo approach  to determine the statistical
spread or uncertainty in the magnetic field parameters as well
as the non-potentiality parameters of active region arising due
to polarimetric noise.
The approach is as follows:\\
 (1) First we take real observations of a sunspot by {\it Hinode}
SOT/SP and invert it with a Milne-Eddington code to retrieve magnetic field vector.\\
(2) resolve the azimuthal ambiguity in the transverse field
using acute angle method.\\
(3) consider this vector field as the ``true field" [$B_0$,
$\gamma_0$, $\psi_0$], and the derived quantities
$\alpha(B_0,\gamma_0, \psi_0)$ and
$\widehat{\Psi}(B_0,\gamma_0, \psi_0)$ as ``true
non-potentiality parameters" [$\alpha_0$,$\widehat{\Psi}_0$],
for reference.\\
(4) using the ``true field" value at each pixel we generate
synthetic Stokes profiles and add polarimetric noise of level
0.5\% of continuum intensity. These profiles are then inverted
again under ME approximation to derive field vector [$B$,
$\gamma$, $\psi$] and [$\alpha_g$, $\widehat{\Psi}$]. This
process is repeated 100 times, each time
 with different realization of polarimetric noise. The  uncertainty  in the field
 strength, inclination and azimuth for each pixel and the uncertainty in the parameters $\alpha_g$ and
$\widehat{\Psi}$ for the vector map is then estimated from the
1-$\sigma$ standard deviation in the 100 values of these
parameters. This standard deviation is called the Monte-Carlo
error.

This simulation gives us an idea about the spread (standard
deviation) in derived field vector and the non-potentiality
parameters of a real sunspot, arising due to polarimetric
noise. This knowledge of standard deviation in the magnetic
non-potentiality parameters $\alpha_g$ and $\widehat{\Psi}$ is
important to determine whether the observed changes in these
parameters, for example in relation to flares, are significant
or not. Such studies are yet to be done and the present work
will establish the level of uncertainties in the parameter
$\widehat{\Psi}$ or $\alpha_g$ due to polarimetric noise in
modern observations, such as from {\it Hinode}.

\subsection{Method of Adding the Noise}
We first generate synthetic Stokes profiles corresponding to
the ``true field" using  Milne-Eddington based Stokes profile
synthesis and inversion code named {\it Helix} \citep{Lagg04}.
To these synthetic profiles we add  normally distributed random
noise with the 3-$\sigma$ level of 0.5\% of continuum intensity
($I_c$). The noise, $N(\lambda)$, is added to the synthetic
Stokes profiles,
$S_{syn}(\lambda)$, as follows:\\
 (1)  First a pseudo random-number sequence  is generated which
is normally distributed with a zero mean and a 3-$\sigma$
standard deviation of given level, say L. In our case L=0.5\%
of $I_c$.
$$ N(\lambda) = \verb"Random_Number"(Seed,L)$$
 (2)  This sequence is then added to the synthetic Stokes profiles to
yield noisy Stokes profile, $S_{noi}(\lambda)$.
$$ S_{noi}(\lambda)=S_{syn}(\lambda)  +  N(\lambda) $$

\begin{figure}
\includegraphics[angle=0,scale=.85]{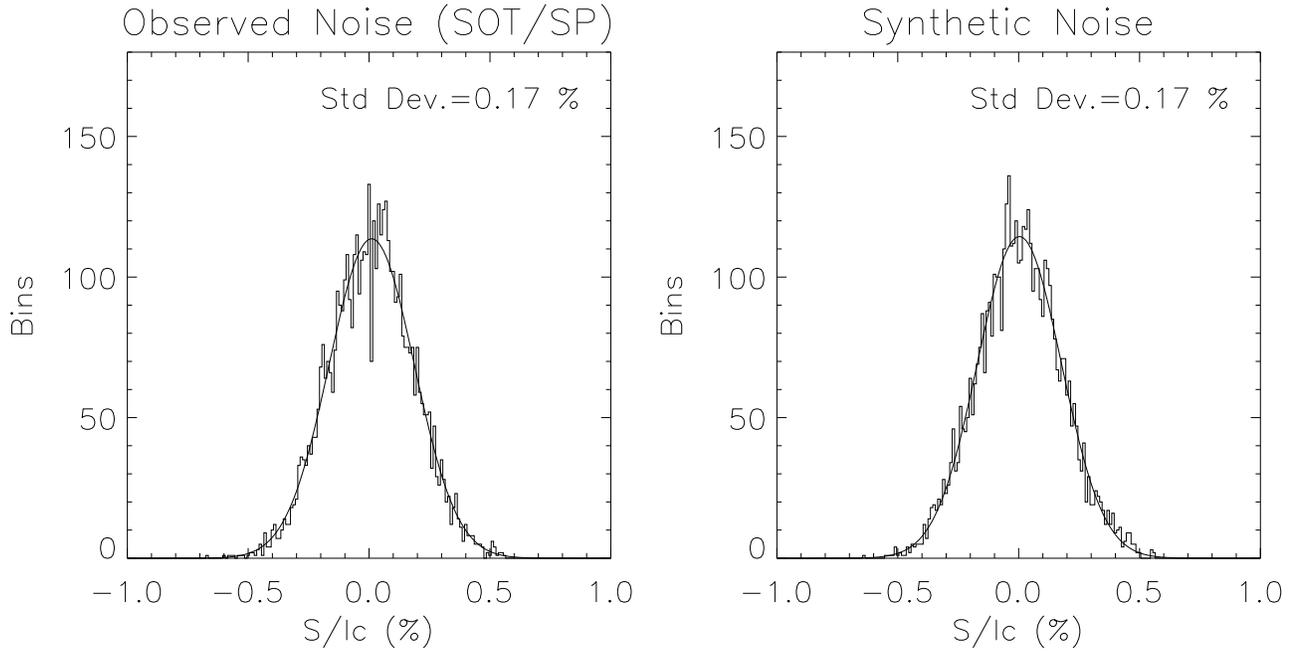}
\caption{ The left panel shows the histogram of the noise in the
Stokes profiles ($S/I_c$) for typical SOT/SP ``fast-mode" scan.
The right panel shows the histogram of normally distributed psuedo-noise
added in the synthetic profiles for Monte-Carlo simulations.
The solid line shows the fitted Gaussian with 1-$\sigma$
noise level is about 0.17\% of $I_c$.}
\label{fig:polnoise}
\end{figure}

 The noise level is estimated from the observed signals in the
continuum of the Stokes spectra. A continuum window between
6302.83 \AA~ and 6303.28 \AA is selected for monitoring the
noise in Stokes signal. The left  panel of the
figure~\ref{fig:polnoise} shows the histogram of noise in the
observed Stokes profiles for a large number of pixels (1024
pixels) for a typical SOT/SP scan in ``fast-map" observing
mode. A Gaussian fit to this distribution yields a 3-$\sigma$
value of 0.5\% of $I_c$. This is the noise level that we used
in our simulations. The histogram in the right panel of
figure~\ref{fig:polnoise} shows the distribution of artificial
noise that we add to the synthetic profiles. These profiles are
then inverted with the {\it Helix} code. This process is
repeated 100 times and so for each pixel we have 100 values
distributed around a mean value.

While modern spectro-polarimetric observations have typically
noise levels of the order of 0.5\% of $I_c$, archived
observations from the ground-based instruments, which might be
used for synoptic studies, may have higher levels of noise.
Therefore, we also carried out an exercise to check the
variation in the magnetic as well as non-potentiality
parameters, $\widehat{\Psi}$ and $\alpha_g$, with increasing
polarimetric noise levels.  We added four different levels of
noise, i.e., 0.01, 0.1, 0.5 and 1 \% of continuum intensity.

\subsection{Method of computing $\alpha_g$ and $\widehat{\Psi}$}
The computation of  the non-potentiality parameter $\alpha_g$
and $\widehat{\Psi}$ is done as proposed by Tiwari {\it et al.}
(2009a, 2009b). The expression for computation of these
parameters is given in equations (1) and (2) in the
introduction section. Only those pixels which have field values
above a certain noise level are analyzed. This filtering is
done in the following way: we select a quiet region on the Sun
and evaluate 1-$\sigma$ deviation in the three vector field
components $B_x$, $B_y$ and $B_z$ separately. The box selected
for this estimation is shown in top-left panel of
figure~\ref{fig:bgammachi}. For transverse vector fields, we
take 1-$\sigma$ noise level as the resultant deviations
obtained in $B_x$ and $B_y$. Only those pixels where transverse
and line-of-sight (LOS) field both are together greater than
twice the above mentioned noise level of 1-$\sigma$ are
analyzed.  The 1-$\sigma$ value for longitudinal and transverse
field inside the box is 380 and 258 Gauss, respectively.

\section{Results}
\subsection{Effect of noise on magnetic field parameters}
\begin{figure}
\includegraphics[angle=0,scale=0.75]{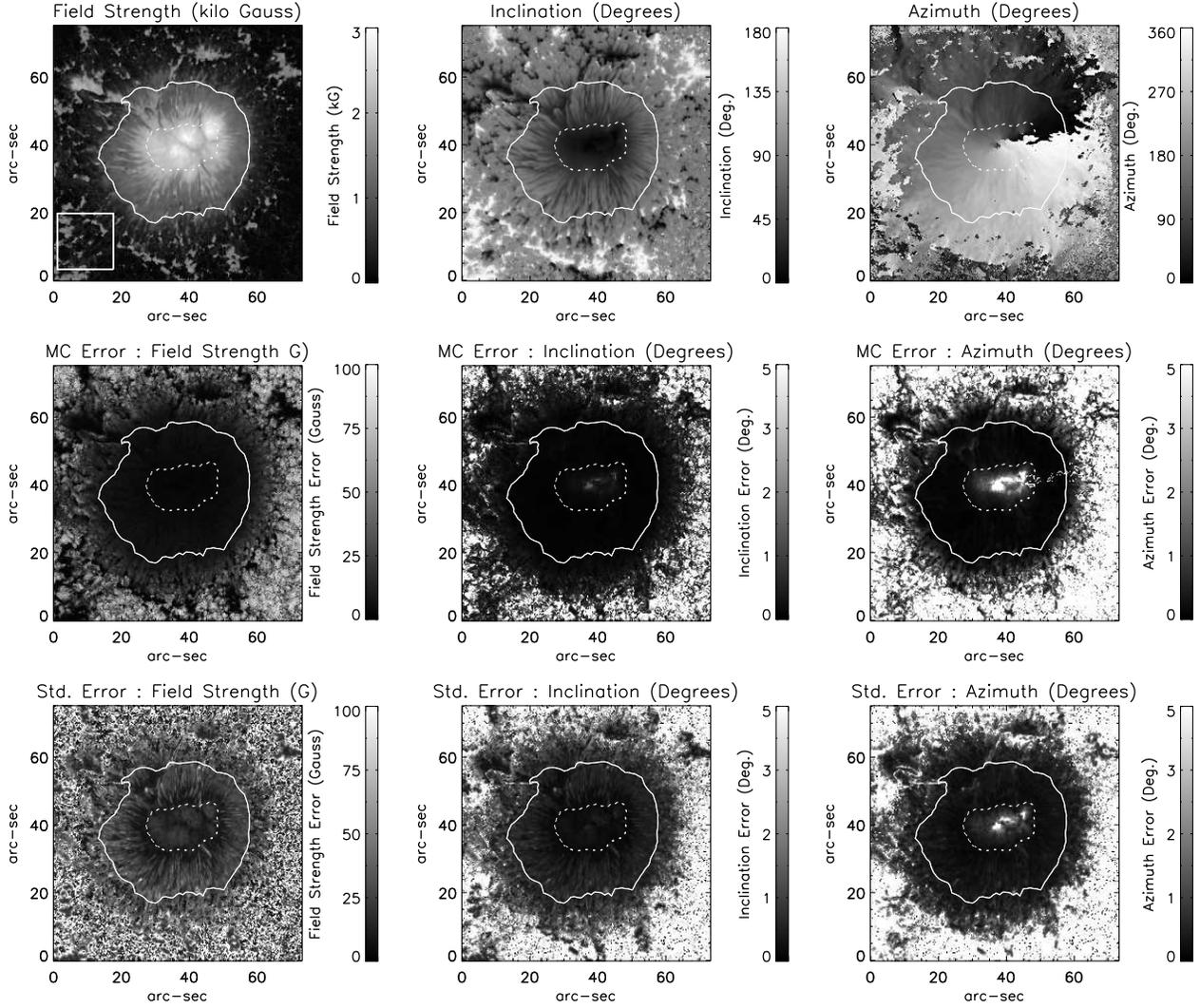}
\caption{The top row shows the  maps of magnetic field strength,
inclination and azimuth for the sunspot in NOAA 10933 derived
using Milne-Eddington inversion. The solid and dashed line
contours mark the penumbral and umbral regions, respectively.
The white box in top left panel marks the box used for
estimation of noise levels in magnetic field components. The
middle row shows the error maps (1-$\sigma$ standard deviation)
for field strength, inclination and azimuth derived using
Monte-Carlo method. The bottom row shows the maps of the
standard error in Milne-Eddington fitting for the field
strength, inclination and azimuth. }
\label{fig:bgammachi}
\end{figure}

The figure~\ref{fig:bgammachi} shows the effect of noise on
magnetic field parameters for the NOAA 10933 sunspot. The top
row shows the initial magnetic field parameters [$B_0$,
$\gamma_0$, $\psi_0$] derived by ME inversion of Stokes
profiles obtained by {\it Hinode} SOT/SP scan during 07:20 UT
on Jan 5, 2009. The middle row shows the map of Monte-Carlo
error, i.e., the 1-$\sigma$ standard deviation in the magnetic
field parameters, derived using the Monte-Carlo method. The
bottom row shows the map of standard error in the field
parameters from the least-squares fit of the Stokes profiles.
The maximum error (taking into account both the standard error
and Monte-Carlo error estimate) inside the sunspot is less than
50 Gauss for the field strength. While for inclination and
azimuth it is less than few degrees. The azimuth errors are
typically largest in the umbral and plage region where the
field is almost vertical and azimuth is not well defined.

\begin{figure}
\includegraphics[angle=0,scale=1.05]{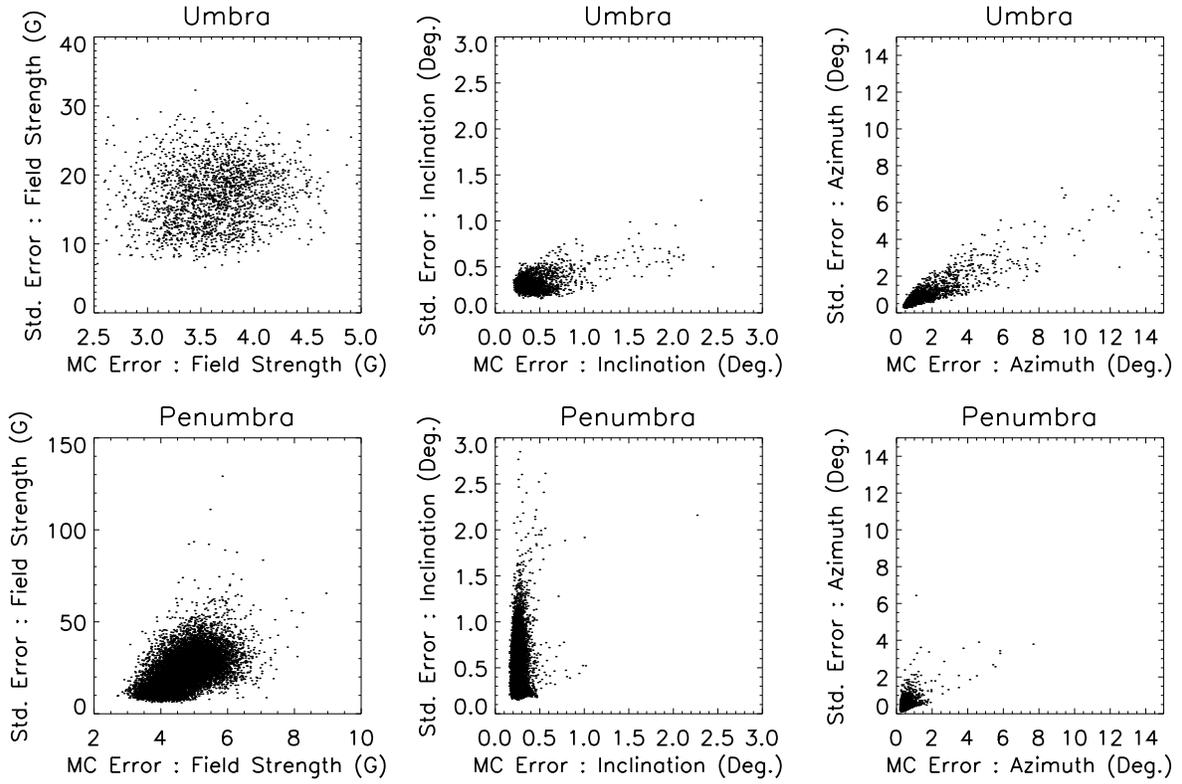}
\caption{Top panel: The scatter plot between the Monte-Carlo errors and
standard errors of fitting are shown for field strength,
inclination and azimuth for all pixels in umbral region. Bottom
panel: The scatter plots for pixels inside penumbral region. }
\label{fig:errors}
\end{figure}

In figure~\ref{fig:bgammachi}, we isolate the umbral and
penumbral regions by using continuum intensity thresholds. We
compare the standard errors and the Monte-Carlo errors in these
two regions in the figure~\ref{fig:errors}. The number of
pixels in the umbral and penumbral regions plotted in
figure~\ref{fig:errors} are 1953 and 13615, respectively. It
may be seen that : \\
(i) in the umbral and penumbral regions the
    Monte-Carlo errors are typically less than 5 Gauss and
    8 Gauss for field strength,  less than 1 and 0.5 degrees
    for field inclination, and less than 5 and 2 degrees
for field azimuth, respectively.\\
 (ii) the standard error is
larger than Monte-Carlo
    error estimates in both the umbral and penumbral
    regions for field strength as was found by \cite{Plaza2001} in case of the inversion of a single profile.
    While for field azimuth and inclination it has an opposite
    relation, except for middle panel of bottom row.\\

The standard error depends upon the sensitivity of the $\chi^2$
to the changes in a particular parameter. It can so happen that
the Stokes profiles might not have significant sensitivity to
the changes in that parameter. This results in a large standard
error. On the other hand, the Monte-Carlo method examines the
changes in the derived parameters resulting from the changes in
the Stokes profiles arising due to random fluctuations  in the
profile as a result of different realizations of noise being
added. Since both errors depend upon different origins of the
profile fluctuations, they need not produce the same results.

In general, considering all the panels in figure 3, it may be
concluded that the standard errors are larger than the
Monte-Carlo errors. Here, one should bear in mind that standard
errors shown in figure 3, are for inversion of real Stokes
profiles which possess asymmetry, apart from the polarimetric
noise. While, Monte-Carlo error correspond to repeated
inversion of purely synthetic, and therefore symmetric (or
antisymmetric for Stokes-V) profiles with polarimetric noise.
Therefore, the main source of difference between the two type
of errors  could be attributed to the presence of Stokes
asymmetry in the real stokes profiles leading to large value of
standard error. Ideally, one would like to perform a
Monte-Carlo error estimate by obtaining several simultaneous
observational datasets of a real sunspot where the Stokes
profiles would have inherent Stokes asymmetry as well as
different realization of polarimetric noise. However, in the
absence of such a possibility we used present method to study
the effect of polarimetric noise alone, neglecting Stokes
asymmetry. The results of the present study are therefore only
the lower limits of the errors that can  occur practically in
real observations as well as in the non-potentiality parameters
$\widehat{\Psi}$ and $\alpha_g$.

\subsection{Effect of noise on Azimuthal Ambiguity Resolution}
 It is well known that the Zeeman effect diagnostics cannot
detect the direction of the transverse field component and so a
180 degree ambiguity remains in the determination of the field
azimuth. Various methods, however, have been developed by the
researchers to resolve this ambiguity using different arguments
(Leka et al. 2009). One of the most common and widely
applicable method is the so-called acute angle method. In this
method the angle between the observed and potential transverse
field, i.e., $\theta = acos(B_o^{obs}.B_t^{pot})/(\mid
B_o^{obs}\mid \mid B_t^{pot}\mid)$, is computed and the
solution for which the value of $\theta$ is an acute angle is
considered as the correct solution. However, in the presence of
polarimetric noise in the observations, this method also can
fail specially in situations where the observed field is highly
sheared. Such highly sheared regions are found near polarity
inversion line (PIL) of the active regions.

Fortunately, in a normal round sunspot, like the one used in
our simulations, there are no high-shear regions  and therefore
the azimuthal ambiguity is easily solved with acute angle
method. Nevertheless, we need to check the effect of
polarimetric noise on the azimuth ambiguity resolution in our
simulations before we examine the uncertainty in the
$\widehat{\Psi}$ and $\alpha_g$ parameter of the sunspot in 100
realizations. After resolving the azimuthal ambiguity using
acute angle method for the 100 realizations of the sunspot
vector maps, we made a map of 1-$\sigma$ standard deviation for
azimuth angle, as shown in middle-right panel of
figure~\ref{fig:bgammachi}. If the ambiguity is not resolved
properly in these 100 realizations, there will be fluctuations
of the order of 180 degrees, leading to a large value of
standard deviation in azimuth value for a given pixel.
However, in the right panel of the middle row in figure~\ref{fig:bgammachi} we see that:\\
(i) the  resolution of the azimuth angle is quite stable for
the most part within the sunspot, specially in the penumbral
region. Outside the sunspot, in the quiet and facular areas,
the errors are large. This is mainly due to poor SNR in these
areas as a result of weaker and/or vertical fields.\\
(ii) As we go towards the umbra the values of standard
deviation are large. We examined the values of the field
azimuth in these regions and found that values do not vary to
the extent of 180 degrees and therefore the large errors in
umbral region is not due to ambiguity solver but due to poor
SNR in Stokes-Q and U observations. Only in the very central
part of the umbra, where the field inclination is close to 90
degrees, the azimuth loses its meaning and so we see a large
spread.

Further, we also checked the effect of ambiguity solver with
increasing noise i.e., four different levels  of noise of 0.01,
0.1, 0.5 and 1 \% of continuum intensity. Here also the azimuth
ambiguity resolution is stable for most part of the sunspot.

\subsection{Effect of noise on $\widehat{\Psi}$ and $\alpha_g$}
\begin{figure}
\includegraphics[angle=0,scale=.85]{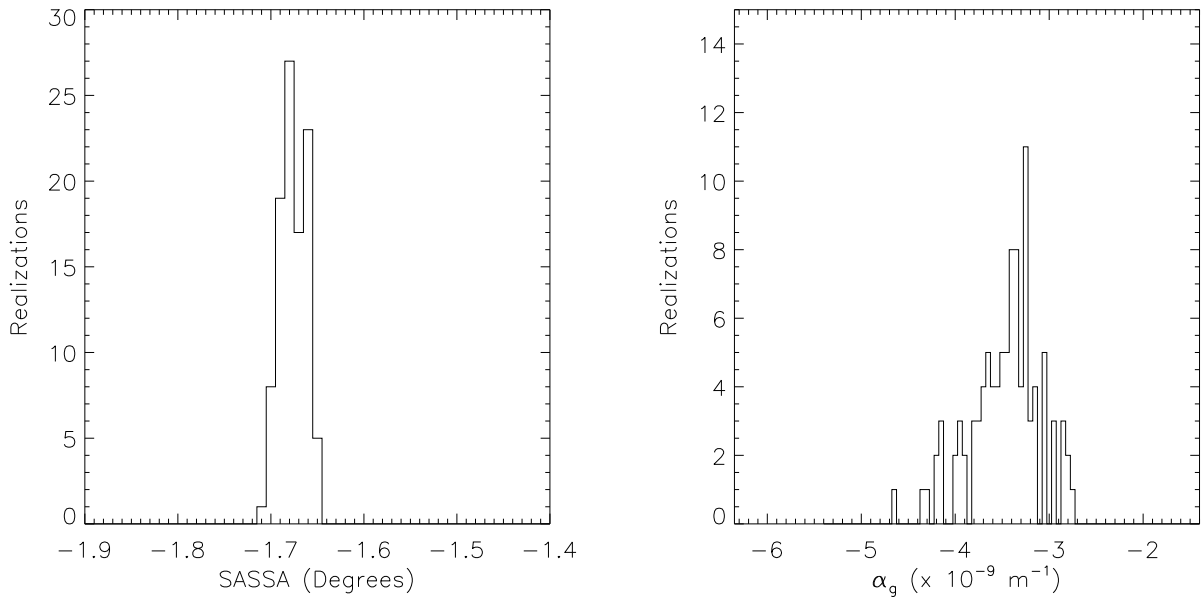}
\caption{
The histograms show distribution  of $\widehat{\Psi}$ and
$\alpha_g$ corresponding to NOAA 10933. The vector maps were
derived hundred times, and each time different realization of
normally distributed noise (3-$\sigma$ = 0.5\% of $I_c$) is
added to the synthetic profiles.   }
\label{fig:histo}
\end{figure}

 The figure~\ref{fig:histo} shows histogram of $\widehat{\Psi}$ and
$\alpha_g$ corresponding to 100 Monte-Carlo realizations of vector field. It can be noticed that:\\
 1. The $\widehat{\Psi}$ is not affected much by noise: The
distribution of $\widehat{\Psi}$ corresponding to 100
realizations show less scatter ($\sim$~1\%) with  $\widehat{\Psi}$ =$ -1.68 \pm 0.014~ ^\circ$.\\
2. The values of $\alpha_g$ are more affected: The distribution
shows large scatter ($\sim$~10\%) in values with $\alpha_g = -3.5 \pm 0.37~ (\times 10^{-9} m^{-1})$.\\
 These results show that the $\widehat{\Psi}$ may be more
reliable as compared to $\alpha_g$ for typical polarimetric
noise present in modern spectro-polarimetric observations as it
is a lower dispersion parameter.

\begin{figure}
\includegraphics[angle=0,scale=.75]{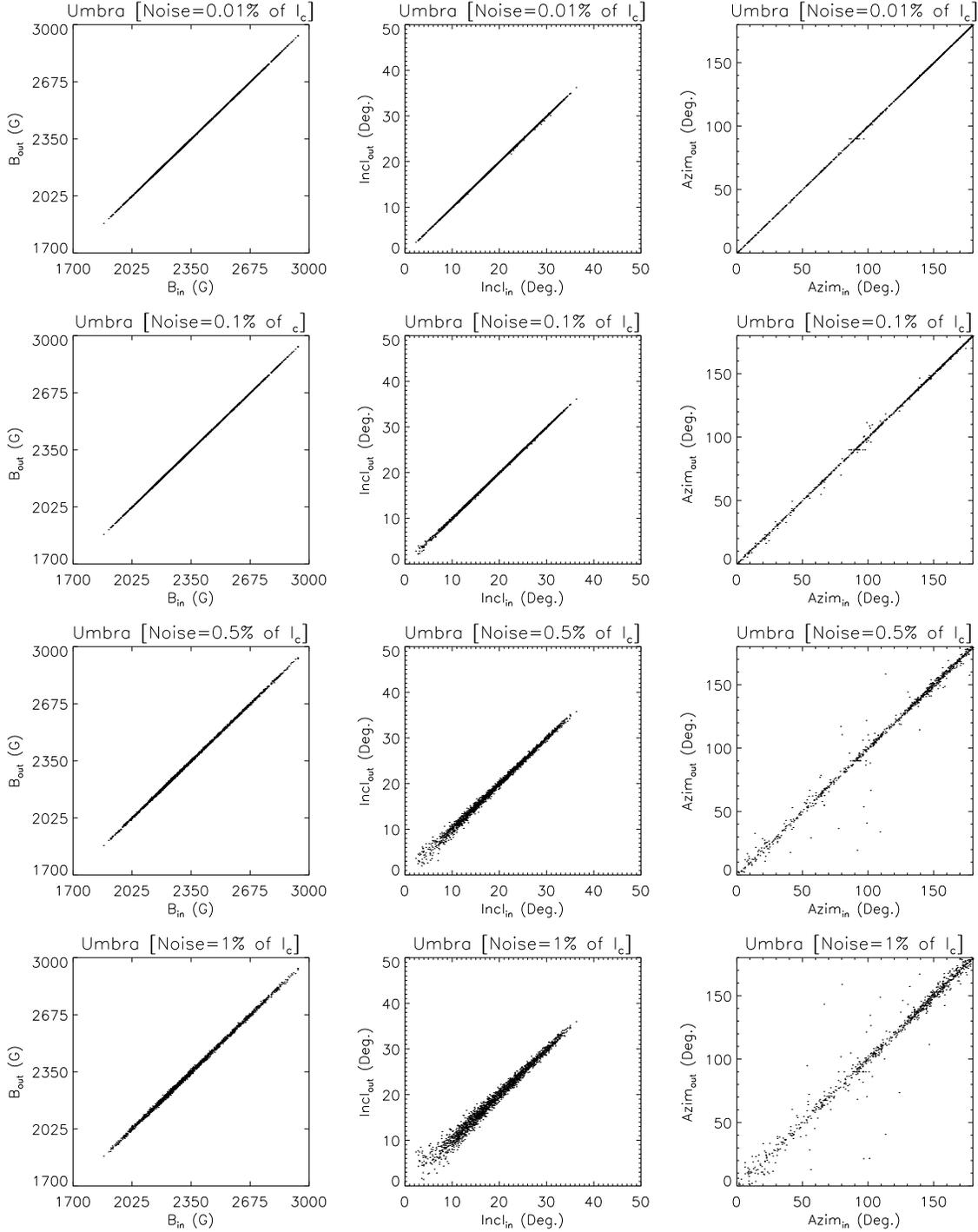}
\caption{The scatter plots between the input and derived magnetic field
parameters after different level of noise is added to Stokes
profiles. The top of each panel shows the 3-$\sigma$ level of
normally distributed noise that is added to the synthetic
profiles. The x-axis shows the input magnetic field parameter
used in the profile synthesis and the y-axis shows the output
value of the parameter derived from the noise added synthetic profile.}
\label{fig:noisescu}
\end{figure}

\begin{figure}
\includegraphics[angle=0,scale=.75]{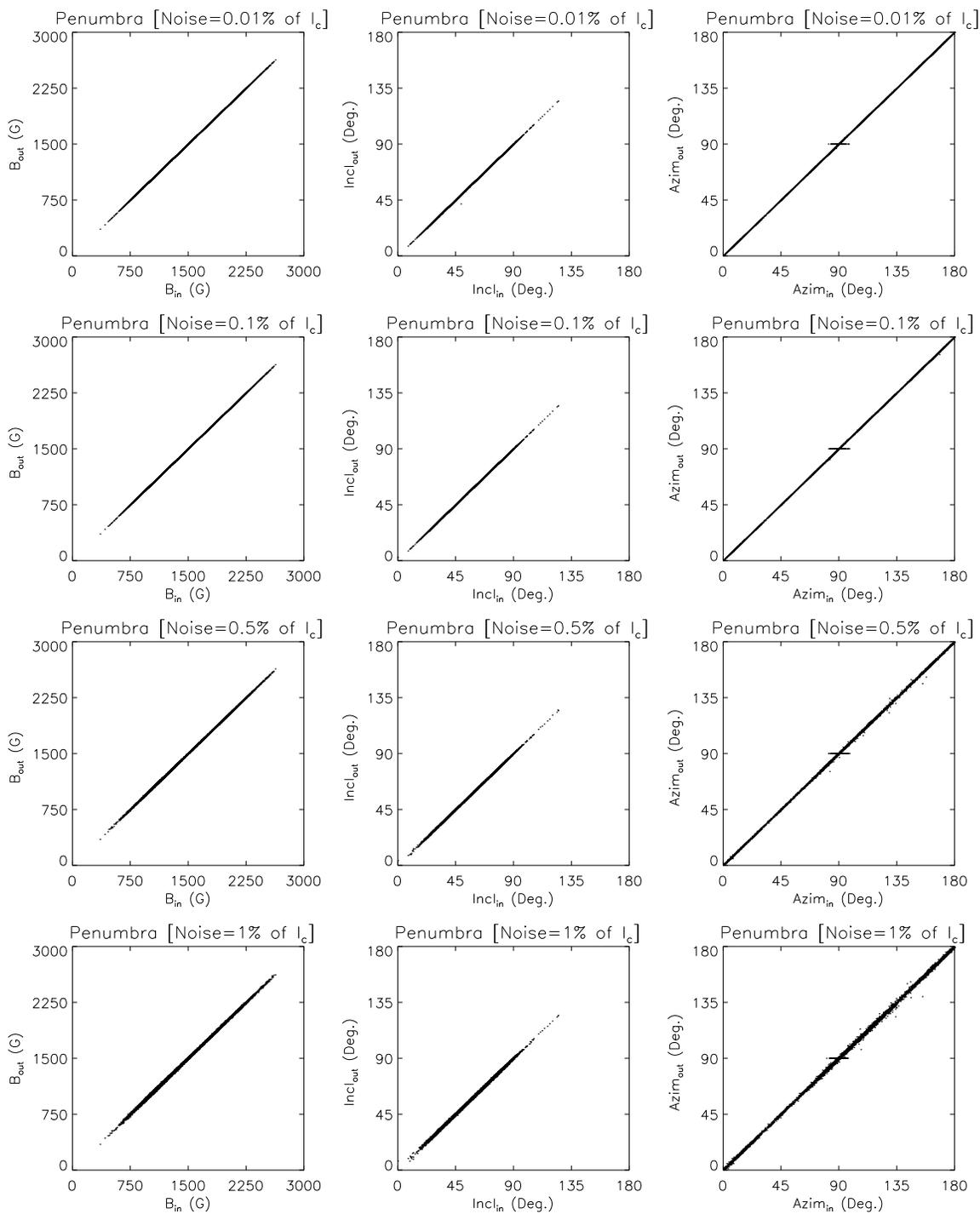}
\caption{Same as figure 7, but for penumbral region.}
\label{fig:noisescp}
\end{figure}

We also check the effect of increasing noise in the
polarimetric measurements on the magnetic as well as
non-potentiality parameters of the same sunspot. In order to
compare the effect of increasing noise on the field parameters
in the sunspot we isolate the umbral and penumbral region, as
was done in figure~\ref{fig:bgammachi}, and do a scatter plot
between the input and output field parameters, for different
levels of noise. The scatter plot for umbral and penumbral
region is shown in figure~\ref{fig:noisescu} and
figure~\ref{fig:noisescp} respectively. It can be seen that, in
the umbra, where the field is mostly vertical, the azimuth
determination is more affected with increasing noise than the
field strength and inclination. Specially,  for 0.5\% and 1\%
noise levels the  spread is large in the azimuth and
inclination values. In comparison, the scatter plot for the
penumbral region, where the field is not so vertical,  shows
that the azimuth is determined with less spread even for 0.5\%
and 1\% noise levels, respectively.


Table~\ref{tab:2} lists the non-potentiality parameters,
derived from the ME inverted magnetic field vector, after
adding noise of different levels in the synthetic Stokes
profiles.  In Table 1, the spatial fluctuations of the shear
angle $\Psi$ as well as local $\alpha$ in the map, are
represented as $\sigma_{\Psi}$ and $\sigma_{\alpha}$. Starting
with no noise and then adding random noise of 0.01, 0.1, 0.5
and 1 \% of  continuum intensity in the Stokes profiles we get
the following results: (i)  The sign of the twist in the
sunspot magnetic field, as inferred by both $\alpha_g$ and
$\widehat{\Psi}$,  is negative and is not affected even when
the SNR of Stokes profiles is poor, (ii) the absolute value of
$\widehat{\Psi}$  tends to decrease systematically with
increasing noise in the Stokes profiles, and (iii) the
fluctuation of 2.5\% in the $\widehat{\Psi}=-1.65\pm0.04^\circ$
is less than the fluctuation of about 5\%  in
$\alpha_g=3.15\pm0.17 (\times10^{-9}m^{-1})$.

\section{Discussion}
 Developing quantitative measures of magnetic non-potentiality
in solar active regions is very important for flare research.
The line-of-sight magnetic field alone is insufficient for this
purpose, therefore people started measuring  vector magnetic
fields. The stresses in the magnetic field were quantified in
terms of the magnetic shear angle, by comparing the observed
field with the potential-field (or current-free fields). The
distribution of shear angle over active region showed that the
filament bearing PIL regions are characterized by high-shear
and are potential sites for flares. Shear, as a local measure
of non-potentiality can be easily extended as a whole active
region measure, like SASSA $\widehat{\Psi}$ (Tiwari et al.
2009b).

Other physical measures of whole active region non-potentiality
are the virial estimate of free-energy and $\alpha_g$. Both of
these measures require the photospheric magnetic field to be
force-free.  However, force-free condition may not be justified
as the plasma $\beta$ in the photosphere is not much smaller
than unity and so the non-magnetic forces are not negligible.
The effect of the polarimetric noise on the virial free-energy
estimate was determined by Klimchuck et al. (1992) and Tiwari
et al. (2009a). However, the results of these studies are valid
for analytic force-free field solution of \citep{Low82}. While,
in real sunspots, the effect of non-force-freeness of
photospheric field on virial energy and $\alpha_g$ may be
larger than the effect of polarimetric errors quantified by
Klimchuck et al. (1992) and Tiwari et al. (2009a). The whole
active region non-potentiality characterized by SASSA
$\widehat{\Psi}$ is, however,  free of force-free assumption
and in the present work we studied the effect of polarimetric
noise on this parameter.

We simulated polarimetric profiles with noise level of 0.5\% of
$I_c$, which is typical of {\it Hinode} SOT/SP observations as
shown in figure 1. It must be noted that the synthesis of these
profiles is done under the assumption of ME model atmosphere
and so the present results pertain to symmetric Stokes profiles
only. The asymmetry of the Stokes profiles would lead to a
systematic rather than random effect which needs to be
quantified in a separate study. In hundred realizations of such
noisy symmetric Stokes profiles we found that the parameter
$\widehat{\Psi}$ is statistically more stable than $\alpha_g$.
The reason for this stability of $\widehat{\Psi}$ as compared
to $\alpha_g$ can be understood as follows. In Eqn. (2)
$\widehat{\Psi}$ is derived as a simple summation of angles.
So, the random errors would cancel each other in summation.
While, in Eqn. (1) we note that $\alpha_g$ depends upon three
components $B_x$, $B_y$ and $B_z$ and is not a simple summation
and therefore the random errors in the three parameters would
not vanish statistically. Thus, $\alpha_g$ would be more
vulnerable to the noise as compared to $\widehat{\Psi}$.

Further, in the presence of different noise levels i.e., 0.01,
0.1, 0.5 and 1\% of $I_c$, it is found that the standard
deviation in $\widehat{\Psi}$ is less vulnerable to increasing
noise than $\alpha_g$. The absolute value of $\widehat{\Psi}$
tends to reduce systematically with increasing noise.

Further, one should keep in mind that the accuracy of derived
magnetic field parameters also depends upon the number of free
parameters (parameters of the model atmosphere) used in the
fitting procedure. The present results pertain to a
single-component model atmosphere with stray-light component.
The results would be different when one employs more number of
free parameters, like two-component model atmosphere (Leka
2001; Lites et al. 2002) or depth-dependent (stratified) model
atmosphere used in SIR inversions. In general the uncertainty
in the best-fit model parameters will deteriorate with
increasing number of model (or free) parameters.

\section{Conclusion}
We generated  an ensemble of artificial spectro-polarimetric
dataset for a given sunspot. This ensemble of data is then
inverted to give an ensemble of vector maps. These maps are
then used to estimate of uncertainties in the field parameters
as well as  in the non-potentiality parameters, $\alpha_g$ and
$\widehat{\Psi}$, of the sunspot. The standard errors of ME
fitting are given by the inversion codes according to
\cite{Press86} and \cite{Bellot2000}. The Monte-Carlo errors
give us an independent method to cross-check the standard error
estimates.  In figure 3, we show that the  Monte-Carlo errors
for field strength, inclination and azimuth in the sunspot is
determined within $\pm$ 8 G, $\pm$ 1 degree and $\pm$ 5 degree,
respectively. A comparison shows that the Monte-Carlo errors in
field strength are typically smaller than the standard errors,
while it is opposite for the field azimuth and inclination,
except for the field inclination in penumbra where again the
Monte-Carlo errors are less than standard errors.  In general,
the standard errors are more conservative estimates because
they include the effects of polarimetric noise as well as those
of Stokes asymmetry, while Monte-Carlo errors account only for
the polarimetric noise.

The effect of polarimetric noise on the  parameters
characterizing the non-potentiality of a sunspot magnetic field
suggests that  $\widehat{\Psi}$ is more robust than $\alpha_g$
 (as shown by the histogram in figure 4). Further,
$\widehat{\Psi}$ appears to be a stable parameter with
increasing noise in the polarimetric data  (Table 1).

 Thus, $\widehat{\Psi}$
will be useful for studying the evolution of non-potentiality
of the active region magnetic fields, which in turn, can help
in the prediction of flare occurrence. The parameter
$\widehat{\Psi}$ in a large number of active regions as well as
the evolution of $\widehat{\Psi}$ in a flaring regions will be
evaluated in a future work.

\acknowledgments We thank the referee for detailed comments and
suggestions. We thank Dr. Bruce Lites for helping in the
estimation of the standard errors. We also would like to thank
Dr. Andreas Lagg for providing his HeliX code used in this
study.  MERLIN inversions of {\it Hinode} SOT/SP were conducted
at NCAR under the framework of the Community
Spectro-polarimtetric Analysis Center (CSAC). {\it Hinode} is a
Japanese mission developed and launched by ISAS/JAXA, with NAOJ
as domestic partner and NASA and STFC (UK) as international
partners. It is operated by these agencies in co-operation with
ESA and NSC (Norway). We thank the IFCPAR (Indo-French Centre
for Promotion of Advanced Research) for the financial support
under Project 3704-1.


\begin{thebibliography}{28}
\expandafter\ifx\csname natexlab\endcsname\relax\def\natexlab#1{#1}\fi

\bibitem[{{Abramenko} {et~al.}(1996){Abramenko}, {Wang}, \&
  {Yurchishin}}]{Abramenko96}
{Abramenko}, V.~I., {Wang}, T., \& {Yurchishin}, V.~B. 1996, \solphys, 168, 75

\bibitem[{{Bao} \& {Zhang}(1998)}]{Bao98}
{Bao}, S., \& {Zhang}, H. 1998, \apjl, 496, L43+

\bibitem[{{Bellot Rubio} {et~al.}(2000){Bellot Rubio}, {Ruiz
    Cobo}, \& {Collados}}]{Bellot2000}
    {Bellot Rubio}, L. R., {Ruiz Cobo}, B. \& {Collados}, M. 2000, \apj,
    535, 475

\bibitem[{{Chatterjee} {et~al.}(2006){Chatterjee}, {Choudhuri}, \&
  {Petrovay}}]{Chatter06}
{Chatterjee}, P., {Choudhuri}, A.~R., \& {Petrovay}, K. 2006, \aap, 449, 781

\bibitem[{{D{\'e}moulin} \& {Pariat}(2009)}]{Demoulin09}
{D{\'e}moulin}, P., \& {Pariat}, E. 2009, Advances in Space Research, 43, 1013

\bibitem[Falconer et al.(2008)]{Falconer2008}
    Falconer,
    D.~A., Moore, R.~L., \& Gary, G.~A.\ 2008, \apj, 689, 1433

\bibitem[{{Gosain} {et~al.}(2009){Gosain}, {Venkatakrishnan}, \&
  {Tiwari}}]{Gosain09}
{Gosain}, S., {Venkatakrishnan}, P., \& {Tiwari}, S.~K. 2009, \apjl, 706, L240

\bibitem[{{Gosain} {et~al.}(2004){Gosain}, {Venkatakrishnan}, \&
  {Venugopalan}}]{Gosain04}
{Gosain}, S., {Venkatakrishnan}, P., \& {Venugopalan}, K. 2004, Experimental
  Astronomy, 18, 31

\bibitem[{{Gosain} {et~al.}(2006){Gosain}, {Venkatakrishnan},
    \& {Venugopalan}}]{Gosain06} {Gosain}, S., {Venkatakrishnan}, P.,
\& {Venugopalan}, K. 2006, Journal of
  Astrophysics and Astronomy, 27, 285

\bibitem[{{Hagino} \& {Sakurai}(2004)}]{Hagino04}
{Hagino}, M., \& {Sakurai}, T. 2004, \pasj, 56, 831

\bibitem[{{Hagino} \& {Sakurai}(2005)}]{Hagino05}
---. 2005, \pasj, 57, 481

\bibitem[{{Harvey}(1969)}]{Harvey69}
{Harvey}, J.~W. 1969, PhD thesis, UNIVERSITY OF COLORADO AT BOULDER.

\bibitem[{{Ichimoto} {et~al.}(2008){Ichimoto}, {Lites}, {Elmore}, {Suematsu},
  {Tsuneta}, {Katsukawa}, {Shimizu}, {Shine}, {Tarbell}, {Title}, {Kiyohara},
  {Shinoda}, {Card}, {Lecinski}, {Streander}, {Nakagiri}, {Miyashita},
  {Noguchi}, {Hoffmann}, \& {Cruz}}]{Ichimoto08}
{Ichimoto}, K., {et~al.} 2008, \solphys, 249, 233

\bibitem[Klimchuk et al.(1992)]{Klimchuk1992} Klimchuk,
    J.~A., Canfield, R.~C., \& Rhoads, J.~E.\ 1992, \apj, 385,
    327

\bibitem[{{Kosugi} {et~al.}(2007){Kosugi}, {Matsuzaki}, {Sakao}, {Shimizu},
  {Sone}, {Tachikawa}, {Hashimoto}, {Minesugi}, {Ohnishi}, {Yamada}, {Tsuneta},
  {Hara}, {Ichimoto}, {Suematsu}, {Shimojo}, {Watanabe}, {Shimada}, {Davis},
  {Hill}, {Owens}, {Title}, {Culhane}, {Harra}, {Doschek}, \&
  {Golub}}]{Kosugi07}
{Kosugi}, T., {et~al.} 2007, \solphys, 243, 3

\bibitem[{{Lagg} {et~al.}(2004){Lagg}, {Woch}, {Krupp}, \& {Solanki}}]{Lagg04}
{Lagg}, A., {Woch}, J., {Krupp}, N., \& {Solanki}, S.~K. 2004, \aap, 414, 1109

\bibitem[Leka(2001)]{Leka2001} Leka, K.~D.\ 2001,
    Advanced Solar Polarimetry -- Theory, Observation, and
    Instrumentation, 236, 571

\bibitem[Lites et al.(2002)]{Lites2002} Lites, B.~W.,
    Socas-Navarro, H., Skumanich, A., \& Shimizu, T.\ 2002,
    \apj, 575, 1131

\bibitem[{{Lites} {et~al.}(2007){Lites}, {Elmore}, {Streander}, {Hoffmann},
  {Tarbell}, {Title}, {Shine}, {Ichimoto}, {Tsuneta}, {Shimizu}, \&
  {Suematsu}}]{Lites07}
{Lites}, B.~W., {et~al.} 2007, in Astronomical Society of the Pacific
  Conference Series, Vol. 369, New Solar Physics with Solar-B Mission, ed.
  {K.~Shibata, S.~Nagata, \& T.~Sakurai}, 55--+

\bibitem[{{Low}(1982)}]{Low82}
{Low}, B.~C. 1982, \solphys, 77, 43

\bibitem[{{Mickey} {et~al.}(1996){Mickey}, {Canfield}, {Labonte}, {Leka},
  {Waterson}, \& {Weber}}]{Mickey96}
{Mickey}, D.~L., {Canfield}, R.~C., {Labonte}, B.~J., {Leka}, K.~D.,
  {Waterson}, M.~F., \& {Weber}, H.~M. 1996, \solphys, 168, 229

\bibitem[{{Nandy}(2006)}]{Nandy06}
{Nandy}, D. 2006, Journal of Geophysical Research (Space Physics), 111, 12

\bibitem[{{Pevtsov} {et~al.}(1994){Pevtsov}, {Canfield}, \&
  {Metcalf}}]{Pevtsov94}
{Pevtsov}, A.~A., {Canfield}, R.~C., \& {Metcalf}, T.~R. 1994, \apjl, 425, L117

\bibitem[{{Pevtsov} {et~al.}(1995){Pevtsov}, {Canfield}, \&
  {Metcalf}}]{Pevtsov95}
---. 1995, \apjl, 440, L109

\bibitem[{{Press} {et~al.}(1986){Press}, {Flannery}, \& {Teukolsky}}]{Press86}
{Press}, W.~H., {Flannery}, B.~P., \& {Teukolsky}, S.~A. 1986, {Numerical
  recipes. The art of scientific computing}, ed. {Press, W.~H., Flannery,
  B.~P., \& Teukolsky, S.~A.}

\bibitem[{{Ruiz Cobo} \& {del Toro Iniesta}(1992)}]{Cobo92}
{Ruiz Cobo}, B., \& {del Toro Iniesta}, J.~C. 1992, \apj, 398, 375

\bibitem[{{Scherrer} \& {SDO/HMI Team}(2002)}]{Scherrer02}
{Scherrer}, P.~H., \& {SDO/HMI Team}. 2002, in Bulletin of the American
  Astronomical Society, Vol.~34, Bulletin of the American Astronomical Society,
  735--+

\bibitem[{{Su} {et~al.}(2009){Su}, {Sakurai}, {Suematsu}, {Hagino}, \&
  {Liu}}]{Su09}
{Su}, J.~T., {Sakurai}, T., {Suematsu}, Y., {Hagino}, M., \& {Liu}, Y. 2009,
  \apjl, 697, L103

\bibitem[{{Tiwari} {et~al.}(2009{\natexlab{a}}){Tiwari}, {Venkatakrishnan},
  {Gosain}, \& {Joshi}}]{Tiwari09a}
{Tiwari}, S.~K., {Venkatakrishnan}, P., {Gosain}, S., \& {Joshi}, J.
  2009{\natexlab{a}}, \apj, 700, 199

\bibitem[{{Tiwari} {et~al.}(2009{\natexlab{b}}){Tiwari}, {Venkatakrishnan}, \&
  {Sankarasubramanian}}]{Tiwari09b}
{Tiwari}, S.~K., {Venkatakrishnan}, P., \& {Sankarasubramanian}, K.
  2009{\natexlab{b}}, \apjl, 702, L133

\bibitem[{{Tsuneta} {et~al.}(2008){Tsuneta}, {Ichimoto}, {Katsukawa}, {Nagata},
  {Otsubo}, {Shimizu}, {Suematsu}, {Nakagiri}, {Noguchi}, {Tarbell}, {Title},
  {Shine}, {Rosenberg}, {Hoffmann}, {Jurcevich}, {Kushner}, {Levay}, {Lites},
  {Elmore}, {Matsushita}, {Kawaguchi}, {Saito}, {Mikami}, {Hill}, \&
  {Owens}}]{Tsuneta08}
{Tsuneta}, S., {et~al.} 2008, \solphys, 249, 167

\bibitem[{{Venkatakrishnan} \& {Tiwari}(2009)}]{pvk09}
{Venkatakrishnan}, P., \& {Tiwari}, S.~K. 2009, \apjl, 706, L114

\bibitem[{{Westendorp Plaza} {et~al.}(2001){Westendorp Plaza},
    {del Toro
  Iniesta}, {Ruiz Cobo}, {Martinez Pillet}, {Lites}, \& {Skumanich}}]{Plaza2001}
{Westendorp Plaza}, C., {del Toro Iniesta}, J.~C., {Ruiz Cobo}, B., {Martinez
  Pillet}, V., {Lites}, B.~W., \& {Skumanich}, A. 2001, \apj, 547,
  1130

\end{thebibliography}

%

%



\clearpage


\begin{table}
\caption{Effect of increasing polarimetric noise in the
estimation of $\widehat{\Psi}$ and $\alpha_g$} \label{tab:2}
\begin{tabular}{lcccccr}
\hline
Noise (\% of $I_c$)  & $\widehat{\Psi}$ ($^\circ$) & &$\sigma_{\Psi}$ ($^\circ$)& & $\alpha_g (m^{-1})$ & $\sigma_{\alpha} (m^{-1})$\\
\hline
0 &  -1.692  & & 15.724  & & $-3.334\times10^{-9}$  & $3.572 \times10^{-7}$ \\
0.01 &  -1.688  & & 15.726  & & $-3.360\times10^{-9}$  & $3.573 \times10^{-7}$\\
0.1 &   -1.685 & & 15.702  & & $-3.221\times10^{-9}$  & $3.574 \times10^{-7}$\\
0.5 &   -1.648 & & 15.636  & & $-2.976\times10^{-9}$  & $3.562 \times10^{-7}$\\
1&   -1.604 & & 15.364  & & $-3.026\times10^{-9}$  & $3.539 \times10^{-7}$\\
\hline
\end{tabular}
\end{table}

\end{document}